\documentclass[twocolumn,fp]{jpsj3}
\usepackage{graphicx}% Include figure files
\usepackage{dcolumn}% Align table columns on decimal point
\usepackage{bm}% bold math
\usepackage{color}% color
\usepackage{ulem}% torikeshisen
\usepackage{mathrsfs}% hanamoji 
\usepackage{txfonts}
\def\vector#1{{\boldsymbol{#1}}}

\def\vk{{\vector k}}

\def\vQ{{\vector Q}}

\def\kB{{k_{\rm B}}}

\def\eq.#1{Eq.~(\ref{#1})}
\def\eqs.#1{Eqs.~(\ref{#1})}

\def\refeq#1{(\ref{#1})}
\def\pardif#1#2{\frac{\partial #1}{\partial #2}}

\def\Hc2{{H_{\rm c2}}}
\def\difHc2{{H'_{\rm c2}}}

\newcommand\Equation[2]{
\begin{equation}\label{#1} %%%%% \tag{#??} 
#2
\end{equation}
}
\newcommand\Equationnoeqn[2]{
\begin{equation*}\label{#1} %%%%% \tag{#??} 
#2
\end{equation*}
}
\title{
Stability of Mixed-Symmetry Superconducting States with Broken \\ 
Time-Reversal Symmetry against Lattice Distortions 
}
\author{Hiroshi Shimahara}
\inst{
Graduate School of Advanced Science and Engineering, 
Hiroshima University, \\ 
Higashi-Hiroshima 739-8530, Japan
}

\recdate{November 11, 2020}
\abst{
We examine the stability of mixed-symmetry superconducting states 
with broken time-reversal symmetry 
in 
spatial-symmetry-broken systems, including chiral states, 
on the basis of the free-energy functional 
derived in the weak-coupling theory. 
We consider a generic $\alpha_1 + i \alpha_2$ wave state, 
with $\alpha_1$ and $\alpha_2$ being different symmetry indices 
such as 
$(\alpha_1,\alpha_2) = ({\rm d},{\rm s})$, 
$({\rm p}_x,{\rm p}_y)$, and $({\rm d},{\rm d}')$. 
The time-reversal symmetry of the mixed-symmetry state 
with the $\alpha_1$- and $\alpha_2$-wave components 
is broken when the phases of these components differ, 
and such a state is called 
the time-reversal-symmetry breaking (TRSB) state. 
However, their phases are equated by Cooper-pair scattering 
between these components if it occurs; 
i.e., when the off-diagonal elements 
$S_{\alpha_1 \alpha_2} = S_{\alpha_2 \alpha_1}$ 
of the scattering matrix are nonzero, 
they destabilize the TRSB state. 
Hence, it has often been believed that the TRSB state is stable 
only in systems with a spatial symmetry that guarantees 
$S_{\alpha_1 \alpha_2}=0$. 
We note that, contrary to this belief, 
the TRSB state can remain stable in systems without the spatial symmetry 
when the relative phase shifts so that $S_{\alpha_1\alpha_2} = 0$ is restored, 
which results in 
a distorted TRSB $(\alpha_1 + \alpha_2) + i \alpha_2$ wave state. 
Here, note that the restoration of $S_{\alpha_1\alpha_2} = 0$ 
does not imply that the symmetry of 
the quasi-particle energy $E_{\vk}$ is recovered. 
This study shows that such stabilization of the TRSB state 
occurs when the distortion is sufficiently small 
and $\Delta_{\alpha_1} \Delta_{\alpha_2}$ is sufficiently large, 
where 
$\Delta_{\alpha}$ is the amplitude of the $\alpha$-wave component in 
the TRSB state 
in the absence of the distortion. 
We clarify the manner in which the shift in the relative phase 
eliminates 
$S_{\alpha_1 \alpha_2}$ 
and prove that such a state yields a free-energy minimum. 
We also propose a formula for the upper bound of 
the degree of lattice distortion, 
below which the TRSB state can be stable. 
}

\begin{document}
%% JPSJ %% 
\sloppy
%% JPSJ %% 
\maketitle

\section{\label{sec:introduction}
Introduction 
}

Mixed-symmetry superconducting states 
with broken time-reversal symmetry, including chiral states, 
have been examined by many authors 
as candidate states in exotic superconductors, 
such as cuprate and ruthenate superconductors~\cite{Tsu00,Mac03}. 
The order parameter of the time-reversal-symmetry breaking (TRSB) 
$\alpha_1 + i \alpha_2$ wave state 
is expressed as 
\Equation{eq:chiralDelta}
{
     \Delta_{\vk} 
      = \Delta_{\alpha_1} \gamma_{\alpha_1}(\vk) 
        +  i \Delta_{\alpha_2} \gamma_{\alpha_2}(\vk) , 
     }
where $\alpha_1$ and $\alpha_2$ are symmetry indices 
and $\Delta_{\alpha_1}$ and $\Delta_{\alpha_2}$ are real and nonzero. 
The functions $\gamma_{\alpha}(\vk)$ are principal basis functions 
of symmetries $\alpha$ 
and are assumed to be real and orthonormal. 
In $\Delta_{\vk}$, the TRSB state 
is characterized by the nonzero relative phase 
on the right-hand side of \eq.{eq:chiralDelta}. 
The most established example of 
the chiral TRSB state is 
the Anderson--Brinkman--Morel (ABM) state 
in superfluid $^3$He,~\cite{And61,And73} 
in which $(\alpha_1,\alpha_2) = ({\rm p}_x,{\rm p}_y)$ 
and $\Delta_{{\rm p}_x} = \Delta_{{\rm p}_y}$. 
The chiral and TRSB states have been examined 
as candidate exotic superconducting states in crystals as well. 
For the ruthenate superconductor ${\rm Sr_2RuO_4}$, 
a triplet chiral state analogous to the ABM state has been 
examined~\cite{Mac03,NoteSRONMR}. 
For cuprate superconductors, 
singlet TRSB states with $(\alpha_1, \alpha_2) = ({\rm d},{\rm s})$ 
have been examined~\cite{Joy90,Don95,Mus96,Jur99,Tim19,Liu97,Gho99}. 
For d-wave superconductors on hexagonal crystal lattices, 
such as a doped graphene and ${\rm SrPtAs}$~\cite{Fis14}, 
chiral  ${\rm d} + i {\rm d}'$ wave states have been examined. 
In most studies, the TRSB states have been examined in systems 
with particular crystal-lattice symmetries 
that prohibit Cooper-pair scattering between 
$\alpha_1$- and $\alpha_2$- components. 
Presumably, the reason is that in the presence of this scattering, 
the TRSB states are unstable 
because the scattering equates the phases of the two components; 
hence, no nonzero relative phase is sustained.

\def\SinGE{{\bar S}}

This can be explained using simple equations. 
The TRSB state expressed in \eq.{eq:chiralDelta} satisfies 
the gap equation written in the form 
\Equationnoeqn{eq:BCSgapeq}
{
     \left (
     \begin{array}{c}
     \Delta_{\alpha_1} \\
     i \Delta_{\alpha_2} 
     \end{array}
     \right ) 
     = 
     \left (
     \begin{array}{cc}
       \SinGE_{\alpha_1 \alpha_1} 
     & \SinGE_{\alpha_1 \alpha_2} \\ 
       \SinGE_{\alpha_2 \alpha_1} 
     & \SinGE_{\alpha_2 \alpha_2} 
     \end{array}
     \right ) 
     \left (
     \begin{array}{c}
     \Delta_{\alpha_1} \\
     i \Delta_{\alpha_2} 
     \end{array}
     \right ) . 
     }
The matrix on the right-hand side has a physical meaning similar 
to that of the scattering matrix ${\hat S}$,~\cite{NoteScalS}
which originates from pairing interactions, 
and the elements ${\bar S}_{\alpha\alpha'}$ are real. 
By separating the real and imaginary parts, 
we obtain 
$\SinGE_{\alpha_1 \alpha_2} \Delta_{\alpha_2} 
 = \SinGE_{\alpha_2\alpha_1} \Delta_{\alpha_1} = 0$, 
which implies $\Delta_{\alpha_1} = \Delta_{\alpha_2} = 0$ 
unless $\SinGE_{\alpha_{\ell} \alpha_{\bar \ell}} = 0$; 
here, we define 
\mbox{${\bar \ell} = 2$ and 1} for 
\mbox{$\ell = 1$ and 2}, respectively. 
%% respectively.~\cite{NoteOffD} 
However, 
when $\SinGE_{\alpha_{\ell} \alpha_{\bar \ell}} = 0$ 
is guaranteed by a symmetry of the system 
(hereinafter called the relevant symmetry), 
both $\Delta_{\alpha_1}$ and $\Delta_{\alpha_2}$ can be nonzero 
simultaneously. 
When $\SinGE_{\alpha_{\ell} \alpha_{\bar \ell}} = 0$, 
the pairing interaction 
does not directly mix $\Delta_{\alpha_1}$ and $\Delta_{\alpha_2}$; 
however, 
these components affect each other through the diagonal elements 
$\SinGE_{\alpha_{\ell}\alpha_{\ell}}$, which contain both 
$\Delta_{\alpha_1}$ and $\Delta_{\alpha_2}$. 
In the following, we refer to the systems with and without the 
relevant symmetry as symmetric and 
symmetry-broken systems, respectively.

By contrast, when the relevant symmetry is broken 
by an anisotropic pressure or a chemical pressure 
so that 
$\SinGE_{\alpha_{\ell} \alpha_{\bar \ell}} \ne 0$, 
the $\alpha_1$- and $\alpha_2$-wave components are mixed. 
Hence, neither pure $\alpha_1$-wave state nor pure $\alpha_2$-wave state is 
an eigenstate of the linearized gap equation, 
and between the two $\alpha_1 \pm \alpha_2$ wave eigenstates,
the one with the higher transition temperature $T_{\rm c}$ is 
the physical solution of the gap equation 
near $T_{\rm c}$. 
Although this state is not a TRSB state, 
we can consider the TRSB state that has the order parameter 
of the form 
\Equation{eq:chiralphi}
{
     \Delta_{\vk} 
      = \Delta_{\alpha_1} \gamma_{\alpha_1}(\vk) 
        +  i e^{i \phi} \Delta_{\alpha_2} 
           \gamma_{\alpha_2}(\vk) 
     }
at low temperatures,~\cite{NoteSym} 
where we introduced the relative phase 
\mbox{$\phi  \in [-\pi/2,\pi/2]$}. 
We refer to this state as 
the TRSB $(\alpha_1 \pm \alpha_2) + i \alpha_2$ wave state 
%% ~\cite{NoteReIm} 
because \eq.{eq:chiralphi} can be rewritten as 
\Equation{eq:chiralphiReIm} 
{
     \begin{split}
     \Delta_{\vk} = 
     & 
         \Delta_{\alpha_1} \gamma_{\alpha_1}(\vk) 
       - \Delta_{\alpha_2} \sin \phi \, \gamma_{\alpha_2}(\vk)  \\ 
     & ~~~~~ + i \Delta_{\alpha_2} \cos \phi \, \gamma_{\alpha_2}(\vk) . 
     \end{split}
     } 
The states with $\Delta_{\vk}$ 
in \eq.{eq:chiralphi}, i.e., \eq.{eq:chiralphiReIm}, 
include generic $(\alpha_1 \pm \alpha_2) \pm i (\alpha_1 \pm \alpha_2)$ 
wave states because the overall phase factor is arbitrary.~\cite{NoteGnrl}

It may appear evident that 
the $\alpha_1 + i \alpha_2$ wave state distorts into 
the $(\alpha_1 \pm \alpha_2) + i \alpha_2$ wave state 
when the system distorts; 
however, the situation is not so simple, 
because in the distorted system, 
$\SinGE_{\alpha_{\ell} \alpha_{\bar \ell}} = 0$ 
is not guaranteed by the symmetry as in the undistorted system. 
We should note that 
the introduction of the phase factor $e^{i \phi}$ in \eq.{eq:chiralphi} 
does not change the situation that 
the off-diagonal elements 
$\SinGE_{\alpha_{\ell} \alpha_{\bar \ell}}$ equate 
the phases and destabilize the TRSB state 
if they are nonzero. 
The gap equation is modified by $\phi$ as 
\Equation{eq:BCSgapeq_phase}
{
     \left (
     \begin{array}{c}
     \Delta_{\alpha_1} \\
     i e^{i \phi} \Delta_{\alpha_2} 
     \end{array}
     \right ) 
     = 
     \left (
     \begin{array}{cc}
       \SinGE_{\alpha_1 \alpha_1} 
     & \SinGE_{\alpha_1 \alpha_2} \\ 
       \SinGE_{\alpha_2 \alpha_1} 
     & \SinGE_{\alpha_2 \alpha_2} 
     \end{array}
     \right ) 
     \left (
     \begin{array}{c}
     \Delta_{\alpha_1} \\
     i e^{i \phi} \Delta_{\alpha_2} 
     \end{array}
     \right ) , 
     }
and the separation of the real and imaginary parts results 
in   $\cos \phi \, \Delta_{\alpha_2} \SinGE_{\alpha_1\alpha_2} 
    = \cos \phi \, \Delta_{\alpha_1} \SinGE_{\alpha_2\alpha_1} = 0$. 
Hence, unless $\SinGE_{\alpha_{\ell} \alpha_{\bar \ell}} = 0$, 
we obtain 
$\Delta_{\alpha_1} = \Delta_{\alpha_2} = 0$ or 
$\phi = \pm \pi/2$, 
which means that the solutions are not TRSB states.~\cite{NoteDiag} 
Therefore, 
if $\SinGE_{\alpha_{\ell} \alpha_{\bar \ell}} \ne 0$ 
because of a distortion of the lattice, 
the Cooper-pair scattering destabilizes any TRSB states 
between $\alpha_1$- and $\alpha_2$-wave states, 
including $(\alpha_1 \pm \alpha_2) + i \alpha_2$ wave states. 
On the basis of these facts, 
it has often been considered that the TRSB states 
are unstable 
unless the system has the relevant symmetry. 
However, we point out that, 
even in symmetry-broken systems, 
the off-diagonal elements $\SinGE_{\alpha_{\ell} \alpha_{\bar \ell}}$ 
can vanish for an appropriate finite value of $\phi$.

In this paper, we examine the effect of 
the symmetry-breaking distortion of the system 
on the stability of the TRSB state. 
We consider systems that are symmetric in the absence of the distortion 
so that a TRSB state 
is a solution of the BCS weak-coupling gap equation 
at ambient pressure. 
In such systems, 
we can consider the following three possible behaviors a priori: 
(a) The TRSB state is stable only in the symmetric system 
because $\SinGE_{\alpha_{\ell} \alpha_{\bar \ell}} \ne 0$ 
in symmetry-broken systems. 
(b) The TRSB state is actually unstable 
even in the symmetric system 
owing to a fluctuation 
that is not incorporated in the BCS gap equation. 
(c) The TRSB state 
is stable in the symmetry-broken system 
near the symmetric system in the manner mentioned above. 
We should recall that 
even an infinitesimal anisotropic pressure breaks the symmetry. 
Hence, if behavior (a) applies, 
then the symmetric system is at a singular point. 
If it is to be avoided 
because it seems physically implausible, 
either (b) or (c) must apply. 
In the following, we prove that 
behavior (c) applies at low temperatures, 
and as a result, 
the order parameter is distorted into the form 
in \eq.{eq:chiralphiReIm} 
so that $\SinGE_{\alpha_{\ell} \alpha_{\bar \ell}} = 0$ is restored.

The distortion that breaks the relevant symmetry depends on 
$(\alpha_1,\alpha_2)$. 
For 
$(\alpha_1,\alpha_2) = ({\rm d}_{x^2-y^2},{\rm s})$, 
the relevant symmetry is broken by orthorhombic distortion~\cite{Don95,Jur99,Gho99}, 
which is realized by a structural phase transition or 
uniaxial pressures in 
the $[1,0,0]$ and $[0,1,0]$ directions. 
For $(\alpha_1,\alpha_2) = ({\rm p}_x,{\rm p}_y)$, 
the pressures in the $[1,1,0]$ direction break the relevant symmetry. 
By contrast, the uniaxial pressures 
in the $[1,0,0]$,\, $[0,1,0]$, and $[0,0,1]$ directions~\cite{NoteSRO} 
do not break the relevant symmetry in the present sense. 
For the pressures in the former two directions, 
although the state immediately below $T_{\rm c}$ becomes 
either the pure ${\rm p}_x$-wave or pure ${\rm p}_y$-wave state 
depending on the distortion direction, 
distorted chiral ${\rm p}_x \pm i {\rm p}_y$ wave states 
can be solutions of the gap equation 
at sufficiently low temperatures, 
in the sense that they are not destabilized 
by the off-diagonal elements, 
because $S_{{\rm p}_x{\rm p}_y} = S_{{\rm p}_y{\rm p}_x} = 0$ 
is conserved.

For 
$(\alpha_1,\alpha_2) = ({\rm d}_{x^2-y^2},{\rm s})$, 
Jurecka and Schachinger~\cite{Jur99} studied a TRSB state that 
they call the ${\rm s} + i ({\rm s+d}_{x^2-y^2})$ wave state; 
however, it is different from the present distorted TRSB state 
in \eq.{eq:chiralphi}.~\cite{NoteJur99} 
O'Donovan and Carbotte~\cite{Don95} 
studied the TRSB state in a superconductor 
on a two-dimensional orthorhombic lattice 
by a numerical calculation, 
and they found that the distorted TRSB state is 
a solution of the BCS equation in a certain parameter region; 
however, it seems that 
because the free energy was not examined, 
we could not obtain information from their result 
to infer which behavior among (a)--(c) occurs. 
In these previous papers, 
the issue concerning $\SinGE_{\alpha_{\ell} \alpha_{\bar \ell}}$ 
mentioned above is not addressed.~\cite{NoteOrth}

In Sect.~\ref{sec:freeenergy}, 
we derive the free-energy functional in the weak-coupling theory. 
In Sect.~\ref{sec:chiralstate}, 
we examine the stability of the TRSB state 
on the basis of the free-energy functional 
and show that the TRSB state 
with a $\phi$ that eliminates 
$\SinGE_{\alpha_{\ell} \alpha_{\bar \ell}}$ 
yields the free-energy minimum. 
In Sect.~\ref{sec:effectofdistortion}, 
we consider a model in which the lattice distortion 
affects the electron dispersion 
and derive an expression for the shift $\phi$ 
in the relative phase. 
We derive a formula for the upper bound of 
the degree of distortion 
below which the TRSB state can be stable. 
The final section is devoted to a summary and discussion.

\section{\label{sec:freeenergy}
Free-Energy Functional 
}

In this section, we derive the free-energy functional 
${\hat {\cal F}}[\Delta]$ to examine the fluctuation of 
the order parameter $\{ \Delta_{\vk} \! \mid \! {}^\forall\vk \}$ 
of superconductivity. 
The superconductivity induced by Cooper pairs of electrons with spins 
$\sigma$ and ${\bar \sigma} \equiv \pm \sigma$ 
can be examined on the basis of the model Hamiltonian 
$H = H_0 + H_1$ with 
\Equationnoeqn{eq:H0H1}
{
     \begin{split}
     H_0 & = \sum_{\vk,\sigma} \xi_{\vk} 
             c_{\vk \sigma}^{\dagger}
             c_{\vk \sigma} , 
     \\ 
     H_1 & = \frac{1}{2N} 
           \sum_{\vk,\vk'}
           \sum_{\sigma}
           V_{\vk\vk'} 
           [\psi_{\sigma {\bar \sigma}}(\vk')]^{\dagger}
           \psi_{\sigma {\bar \sigma}}(\vk) , 
     \end{split}
     }
where 
$\sigma = + 1$ and $-1$ correspond to 
the up- and down-spins, respectively; 
$\psi_{\sigma{\bar \sigma}} = c_{\vk \sigma} c_{-\vk {\bar \sigma}}$; 
and $N$ denotes the number of lattice sites. 
In the BCS approximation, the free-energy functional 
of the variational parameters 
$\{ E_{\vk} \}$ and $\{ \Delta_{\vk} \}$ 
is expressed as 
\Equation{eq:hatF}
{
     \begin{split} 
     {\hat F}[E,\Delta]  
     = 
     & ~ 
        2 \sum_{\vk} ( E_{\vk} {\cal E}_{\vk} - \xi_{\vk}^2 ) {\cal W}_{\vk} 
        + \sum_{\vk} (\xi_{\vk} - E_{\vk}) 
                                       \\[-2pt] 
     & ~ 
     + \frac{1}{N} \sum_{\vk \vk'} 
       V_{\vk\vk'} {\cal W}_{\vk} {\cal W}_{\vk'} R_{\vk \vk'} 
     \\[-2pt]
     & ~ 
     + \frac{2}{\beta} \sum_{\vk} \ln [ 1 - f(E_{\vk}) ] , 
     \end{split} 
     }
where 
$$
     {\cal W}_{\vk} (E_{\vk}) 
       = \frac{\tanh(\beta E_{\vk}/2)}{2 {\cal E}_{\vk}(\Delta_{\vk}) } , 
     $$ 
${\cal E}_{\vk}(\Delta_{\vk}) 
  = [ \xi_{\vk}^2 + |\Delta_{\vk}|^2 ]^{1/2}$, 
$R_{\vk \vk'} = {\rm Re}[ \Delta_{\vk} \Delta_{\vk'}^{*}]$, 
and 
$f(E) = 1/(e^{\,\beta E} + 1)$. 
The derivation of \eq.{eq:hatF} is outlined in \mbox{Appendix\ref{app:A}} 
with the exact meanings of $\Delta_{\vk}$ and $E_{\vk}$.

For a given set of $\{ \Delta_{\vk} \}$, 
the functional ${\hat F}[E,\Delta]$ is minimum when 
\Equationnoeqn{eq:EDelta}
{
     E_{\vk} = \frac{\xi_{\vk}^2 + D_{\vk}^2}{{\cal E}_{\vk}} 
             \equiv {\hat E}_{\vk}[\Delta] , 
     }
where 
\Equation{eq:Ddef}
{
     D_{\vk}^2 = - \frac{1}{N} \sum_{\vk'} V_{\vk \vk'} 
       {\cal W}_{\vk'} R_{\vk \vk'} . 
     }
Hence, we obtain the reduced form of the free-energy functional 
\Equation{eq:calhatF}
{
     \begin{split}
     {\hat {\cal F}}[\Delta] 
     \equiv & 
       {\hat F}[ {\hat E}[\Delta],\Delta] \\ 
     = & 
         \sum_{\vk} 
         ({\hat E}_{\vk} {\cal E}_{\vk} - \xi_{\vk}^2) 
         {\cal W}_{\vk}({\hat E}_{\vk}) 
       + \sum_{\vk} (\xi_{\vk} - {\hat E}_{\vk}) \\ 
       & ~~~~~ 
       + \frac{2}{\beta} \sum_{\vk} \ln [ 1 - f({\hat E}_{\vk}) ] 
     \end{split}
     }
for the variational parameters $\{ \Delta_{\vk} \}$. 
As shown in \mbox{Appendix\ref{app:A}}, 
the extrema of ${\hat {\cal F}}$ occur when ${\cal E}_{\vk} = E_{\vk}$, 
which results in 
the BCS gap equation for $\Delta_{\vk}$ [i.e., \eq.{eq:BCSeq}].

\section{\label{sec:chiralstate}
Stability of TRSB State in Distorted Systems 
}

The TRSB $\alpha_1 + i \alpha_2$ wave state can be examined 
using the model of pairing interactions with the coupling constant 
\Equationnoeqn{eq:Vkk}
{
     V_{\vk\vk'} = - g_{\alpha_1} \gamma_{\alpha_1}(\vk) \gamma_{\alpha_1}(\vk') 
                   - g_{\alpha_2} \gamma_{\alpha_2}(\vk) \gamma_{\alpha_2}(\vk') , 
     }
where $g_{\alpha_1}$ and $g_{\alpha_2}$ are positive constants. 
We adopt the orthonormal condition 
\Equationnoeqn{eq:gammanorm}
{
     \frac{1}{N} \sum_{\vk} \gamma_{\alpha}(\vk) \gamma_{\alpha'}(\vk) 
       = \delta_{\alpha \alpha'} . 
     }
The distortion of the crystal lattice affects 
both $\xi_{\vk}$ and $V_{\vk\vk'}$. 
However, 
to reproduce the situation of a nonzero 
$\SinGE_{\alpha_{\ell} \alpha_{\bar \ell}}$ destabilizing the TRSB state, 
it is sufficient to retain one of the changes 
in $\xi_{\vk}$ and $V_{\vk \vk'}$ due to the distortion. 
Hence, 
we consider a model in which the relevant symmetry is broken 
in $\xi_{\vk}$ only. 
For example, 
for 
$(\alpha_1,\alpha_2) = ({\rm d}_{x^2-y^2},{\rm s})$, 
the relevant symmetry is broken when 
$\xi(k_x,k_y,k_z) \ne \xi(k_y,k_x,k_z)$. 
For $(\alpha_1,\alpha_2) = ({\rm p}_x,{\rm p}_y)$, 
it is broken 
when 
$\xi(- k_x,k_y,k_z) \ne \xi(k_x,k_y,k_z) \ne \xi(k_x,- k_y,k_z)$, 
where the symmetry 
$\xi(k_x,k_y,k_z) = \xi(- k_x,- k_y,k_z) = \xi(k_y,k_x,k_z)$ 
is kept.

The extremum condition of ${\hat {\cal F}}[\Delta]$ yields 
\Equation{eq:KaaWaa}
{
     \begin{split}
     \SinGE_{\alpha \alpha'} 
      & 
       = g_{\alpha} W_{\alpha \alpha'} , \\
     W_{\alpha \alpha'} 
      & = \frac{1}{N} \sum_{\vk} 
           \gamma_{\alpha}(\vk) W_{\vk} 
           \gamma_{\alpha'}(\vk) , 
           \\
     W_{\vk} 
     & = \frac{\tanh(\beta E_{\vk}/2)}{2 E_{\vk}} , 
     \end{split} 
     }
and $E_{\vk} = [\xi_{\vk}^2 + |\Delta_{\vk}|^2 ]^{1/2}$. 
In these equations, 
if $E_{\vk}$ has the relevant symmetry mentioned above, 
$W_{\vk}$ also has the same symmetry; 
hence, 
$\SinGE_{\alpha_{\ell} \alpha_{\bar \ell}} 
     = g_{\alpha_{\ell}} W_{\alpha_{\ell} \alpha_{\bar \ell}}$ 
vanishes. 
On the other hand, the asymmetry in $E_{\vk}$ 
due to those in $\xi_{\vk}$ and $|\Delta_{\vk}|^2$ 
makes $\SinGE_{\alpha_{\ell} \alpha_{\bar \ell}}$ nonzero 
except for a special case explained below. 
In particular, when $\phi = 0$, 
because $|\Delta_{\vk}|^2$ is symmetric, 
the asymmetry in $\xi_{\vk}$ would result in 
$\SinGE_{\alpha_{\ell} \alpha_{\bar \ell}} \ne 0$; 
hence, the stable state is not 
the TRSB $\alpha_1 + i \alpha_2$ wave state, 
unless the parameters in $\xi_{\vk}$ satisfy an accidental condition 
so that $W_{\alpha_{\ell} \alpha_{\bar \ell}} = 0$. 
By contrast, 
when $\phi \ne 0$, 
a nonzero $\phi$ can eliminate $\SinGE_{\alpha_{\ell} \alpha_{\bar \ell}}$. 
Because both $\xi_{\vk}$ and $|\Delta_{\vk}|^2$ are asymmetric, 
the off-diagonal elements 
$\SinGE_{\alpha_{\ell} \alpha_{\bar \ell}}$ vanish 
if the {\it influence} of the asymmetry in $|\Delta_{\vk}|^2$ 
exactly cancels the {\it influence} 
of the asymmetry in $\xi_{\vk}$ 
{\it in the summation} in \eq.{eq:KaaWaa}. 
Such a cancellation may appear to be accidental and 
difficult to realize; however, in the following, 
we show that it necessarily occurs near the symmetric system 
under a certain condition.

In the subspace of $\{ \Delta_{\vk} \}$ 
in which $\Delta_{\vk}$ has the form presented in \eq.{eq:chiralphi}, 
the functional ${\hat {\cal F}}[\Delta]$ can be regarded as 
a function of $\Delta_{\alpha_1}$, $\Delta_{\alpha_2}$, 
and $\phi$ and thereby denoted as 
${\hat {\cal F}}(\Delta_{\alpha_1}, \Delta_{\alpha_2}, \phi)$. 
Two of the extremum conditions, 
\Equationnoeqn{eq:dFdDeltad}
{
     \frac{\partial {\hat {\cal F}}}{\partial \Delta_{\alpha_1}} 
     = 
     \frac{\partial {\hat {\cal F}}}{\partial \Delta_{\alpha_2}} = 0 , 
     }
yield 
\Equation{eq:newgapeq}
{
     \begin{split}
       {\cal K}_{11} \Delta_{\alpha_1} 
     - {\cal K}_{12} \Delta_{\alpha_2} \sin \phi & = 0 , 
     \\ 
       {\cal K}_{22} \Delta_{\alpha_2} 
     - {\cal K}_{21} \Delta_{\alpha_1} \sin \phi & = 0 , 
     \end{split}
     }
where the matrix elements ${\cal K}_{\alpha\alpha'}$ 
are defined in 
\mbox{Appendix\ref{app:B}}. 
The remaining extremum condition is 
\Equation{eq:dfdphis}
{
     0 = \frac{1}{2N} \frac{\partial {\cal F}}{\partial \phi}
       = \Delta_{\alpha_1} \Delta_{\alpha_2} 
       {\cal K}_{12} 
       \cos \phi . 
     } 
When either $\Delta_{\alpha_1}$ or $\Delta_{\alpha_2}$ 
is equal to zero, 
the solution of \eq.{eq:dfdphis} is a pure state 
and not a TRSB state. 
When $\cos \phi = 0$, 
the solution is an $\alpha_1 \pm \alpha_2$ state, 
which is not a TRSB state either. 
Hence, let us examine the last possibility for 
the solution of \eq.{eq:dfdphis}, that is, 
\Equation{eq:calKdszero}
{
     {\cal K}_{12} = 0 
     }
with both $\Delta_{\alpha_1}$ and $\Delta_{\alpha_2}$ being nonzero. 
Equations~\refeq{eq:newgapeq} and~\refeq{eq:calKdszero} lead to 
\Equation{eq:newgapeq2}
{
       {\cal K}_{11} = 
       {\cal K}_{22} = 0 , 
     }
and \eqs.{eq:calKdszero} and \refeq{eq:newgapeq2} are 
written in the form 
\Equation{eq:LmatrixKvector}
{
     \left ( \!\! 
     \begin{array}{c}
     \, 0 \,\, \\ 
     \, 0 \ \, \\ 
     \, 0 \,\, 
     \end{array} 
     \!\! \right ) 
     = 
     \left ( \!\! 
     \begin{array}{c}
     {\cal K}_{11} \\ {\cal K}_{22} \\ {\cal K}_{12}
     \end{array}
     \!\! \right ) 
     = 
     {\hat {\cal L}} 
     \left ( \!\! 
     \begin{array}{c}
     {\cal K}_{1} \\ {\cal K}_{2} \\ {\cal W}_{12}
     \end{array}
     \!\! \right ) , 
     }
where 
${\cal K}_{\ell}  = 1 - g_{\alpha_{\ell}} {\cal W}_{\ell\ell}$ 
for $\ell = 1,2$ 
and 
\Equationnoeqn{eq:calWcal}
{
     {\cal W}_{k \ell} 
     = \frac{1}{N} \sum_{\vk} 
       {\cal W}_{\vk} 
       \gamma_{\alpha_k}(\vk) \gamma_{\alpha_{\ell}}(\vk) . 
     }
The matrix elements of ${\hat {\cal L}}$ 
are defined in \mbox{Appendix\ref{app:B}}. 
Unless ${\rm det} [ {\hat {\cal L}} ]  = 0$ 
is accidentally satisfied~\cite{NotedetL}, 
\eq.{eq:LmatrixKvector} results in 
\Equation{eq:gapeqresult}
{
     {\cal K}_{1} = {\cal K}_{2} = 0 , 
     ~~~~ {\cal W}_{12} = 0 . 
     }
Equation~\refeq{eq:gapeqresult} reduces \eq.{eq:Ddef} 
to $D_{\vk}^2 = |\Delta_{\vk}|^2$, 
which results in 
$E_{\vk} = {\cal E}_{\vk}$ and $W_{\vk} = {\cal W}_{\vk}$. 
Hence, \eq.{eq:gapeqresult} leads to 
\Equationnoeqn{eq:BCSeqrecovered}
{
     1 = g_{\alpha_1} W_{\alpha_1\alpha_1} 
       = g_{\alpha_2} W_{\alpha_2\alpha_2} 
     }
and 
\Equation{eq:W120}
{
     W_{\alpha_1\alpha_2} = W_{\alpha_2\alpha_1} = 0 , 
     } 
which means that $\SinGE_{\alpha_1 \alpha_2} = \SinGE_{\alpha_2 \alpha_1} = 0$. 
Note that \eq.{eq:W120} is 
neither assumed as an extra condition nor accidental; 
rather, it is derived from the condition of the free-energy extremum. 
Therefore, in distorted systems, 
a finite $\phi$ that gives a free-energy extremum 
restores the condition 
$\SinGE_{\alpha_{\ell} \alpha_{\bar \ell}} = 0$ 
and brings the TRSB state 
back to a solution of the gap equation. 
Indeed, when the lattice distortion is small, for instance, 
\eq.{eq:W120} has a solution for $\phi$: 
\Equation{eq:sinphissol}
{
     \sin \phi = 
     - \frac{W_{\alpha_1\alpha_2}(\phi = 0)}
            {\Delta_{\alpha_1} \Delta_{\alpha_2} 
             W_{\alpha_1\alpha_2}^{(2)}(\phi = 0)} , 
     }
where we define 
\Equation{eq:W2aabb}
{
     W_{\alpha_1\alpha_2}^{(2)} 
     = \frac{1}{N} \sum_{\vk} \frac{W_{\vk}}{E_{\vk}^2} 
       [\gamma_{\alpha_1}(\vk)]^2 
       [\gamma_{\alpha_2}(\vk)]^2 . 
     }
Equation~\refeq{eq:sinphissol} implies that $\phi \ne 0$ 
in symmetry-broken systems 
in which $W_{\alpha_1 \alpha_2}(\phi = 0) \ne 0$. 
The resulting state is a distorted TRSB 
$(\alpha_1 \pm \alpha_2) + i \alpha_2$ wave state with 
an order parameter of the form 
given in \eqs.{eq:chiralphi} and \refeq{eq:chiralphiReIm}.

This behavior can be explained as follows. 
For convenience, 
we define $\epsilon$ as the degree of the distortion from the symmetric system 
in which $W_{\alpha_1\alpha_2}(\phi = 0) = 0$. 
The function $W_{\vk}$ is a functional of 
\Equation{eq:EvkReIm}
{
     E_{\vk} = 
       \Bigl [ 
         \xi_{\vk}^2 + {\bar \Delta}_{\vk}^2 
%%        - 2 \sin \phi \, \Delta_{\alpha_1} \Delta_{\alpha_2} 
%%            \gamma_{\alpha_1}(\vk) \gamma_{\alpha_2}(\vk) 
       - 2 {\cal A}_{\vk} 
       \Bigr ]^{1/2} , 
     }
where ${\bar \Delta}_{\vk} = 
       [   \Delta_{\alpha_1}^2 \gamma_{\alpha_1}^2 
         + \Delta_{\alpha_2}^2 \gamma_{\alpha_2}^2 ]^{1/2}$ 
and 
\Equation{eq:Akphidef}
{
     {\cal A}_{\vk}(\phi) 
        = \sin \phi \, \Delta_{\alpha_1} \Delta_{\alpha_2} 
          \gamma_{\alpha_1}(\vk) \gamma_{\alpha_2}(\vk) . 
     }
Therefore, 
since ${\bar \Delta}_{\vk}$ is symmetric, 
$W_{\vk}$ can be asymmetric because of 
the asymmetry in $\xi_{\vk}^2$ due to $\epsilon \ne 0$ 
and that in ${\cal A}_{\vk}(\phi)$ due to $\phi \ne 0$. 
Because of the asymmetry in $W_{\vk}$, 
it may appear plausible to assume $W_{\alpha_1 \alpha_2} \ne 0$; 
however, in actuality, 
the asymmetry in ${\cal A}_{\vk}(\phi)$ 
%% $- 2 \sin \phi \, 
%% \Delta_{\alpha_1} \Delta_{\alpha_2} 
%%               \gamma_{\alpha_1} \gamma_{\alpha_2}$ 
exactly compensates for 
the {\it influence} of the asymmetry in $\xi_{\vk}^2$ 
{\it in the summation} over $\vk$ in \eq.{eq:KaaWaa}, 
so that $W_{\alpha_1\alpha_2}$ vanishes. 
This does not imply that the asymmetric parts of 
${\cal A}_{\vk}$ and $\xi_{\vk}^2$ directly cancel out, 
making $E_{\vk}$ symmetric. 
In practice, 
because the form of $\xi_{\vk}^2$ is not related to 
that of the gap function, 
this direct cancellation does not occur; 
hence, $E_{\vk}$ remains asymmetric. 
(The symmetry of $E_{\vk}$ is not necessary 
for $W_{\alpha_1\alpha_2}$ to vanish.) 
It also follows from the form of ${\cal A}_{\vk}(\phi)$ 
in \eq.{eq:Akphidef} that if this cancellation is to occur, 
$\Delta_{\alpha_1}\Delta_{\alpha_2}$ must be sufficiently large. 
Conversely, 
when $\Delta_{\alpha_1}\Delta_{\alpha_2}$ is small, 
for example, near the second-order transition temperature, 
the TRSB state is unstable.

When $\epsilon$ is sufficiently small 
and $\Delta_{\alpha_1}\Delta_{\alpha_2}$ is sufficiently large, 
an appropriate finite $\phi \ne 0$ satisfies \eq.{eq:W120} 
as shown in \eq.{eq:sinphissol}. 
Such a value of $\phi$ yields an extremum of 
${\hat {\cal F}}$ as mentioned above. 
Next, we prove that this extremum is a minimum 
for any sufficiently small $\epsilon$, 
considering fluctuations around the extremum. 
Because it is evident that amplitude fluctuations increase 
the free energy (i.e., 
$\partial^2 {\cal F}/\partial \Delta_{\alpha_{\ell}}^2 > 0$), 
we examine phase fluctuations. 
It can be derived that 
\Equationnoeqn{eq:d2Fdphis2}
{
     \frac{1}{2N} \frac{\partial^2 {\hat {\cal F}}}{\partial \phi^2} 
     = \Delta_{\alpha_1} \Delta_{\alpha_2} 
       \left [ \pardif{ {\cal W}_{12} }{\phi} \right ]_0 
       \Gamma_0 + {\rm O}(\epsilon) 
     }
with 
\Equationnoeqn{eq:Gamma}
{
     \Gamma_0 
       = \left [ 
           1 - (g_{\alpha_2} \Delta_{\alpha_1}^2 
              + g_{\alpha_1} \Delta_{\alpha_2}^2) 
                     W_{\alpha_1 \alpha_2}^{(2)} 
         \right ]_0 
     }
at the extremum, where $[\cdots]_0$ indicates the value at $\epsilon = 0$. 
At the extremum, 
$$
     \Delta_{\alpha_1} \Delta_{\alpha_2} 
       \left [ \pardif{ {\cal W}_{12} }{\phi} \right ]_0 
     = 
       \Delta_{\alpha_1}^2 \Delta_{\alpha_2}^2 \cos \phi
       \left [ W_{\alpha_1\alpha_2}^{(2)} \right ]_0 
     > 0 , 
     $$
and $\Gamma_0 > 0$ as proved in \mbox{Appendix\ref{app:C}}. 
Therefore, we obtain 
\Equationnoeqn{eq:d2Fdphi2positive}
{
     \frac{\partial^2 {\hat {\cal F}}}
          {\partial \phi^2 }    > 0 
     }
on the order of $\epsilon^{0}$.
When $\Delta_{\alpha_1} \Delta_{\alpha_2}$ is finite, 
an arbitrary, sufficiently small $\epsilon$ 
does not reverse the sign of 
${\partial^2 {\hat {\cal F}}}/{\partial \phi^2 }$, 
because the value of 
$[{\partial^2 {\hat {\cal F}}}/{\partial \phi^2 }]_{\epsilon = 0}$ 
is finite. 
Hence, 
in a finite region of $\epsilon$ around $\epsilon = 0$, 
the TRSB state (distorted or undistorted depending 
on the value of $\epsilon$) 
yields the free-energy minimum at sufficiently low temperatures.

In the above, it is shown that the TRSB state 
is stable against small fluctuations, 
which implies that the free energy of the TRSB state is 
at a minimum; however, it can be a local minimum. 
To confirm the stability of the TRSB state, 
we must compare its free energy 
with those of the mixed ${\alpha_1} \pm {\alpha_2}$ wave states. 
If the TRSB ${\alpha_1} + i {\alpha_2}$ wave state 
has the lowest free energy at low temperatures 
for $\epsilon = 0$, 
and the differences in the free energy 
between the TRSB and pure states are finite, 
then one of the TRSB $({\alpha_1} \pm {\alpha_2}) + i {\alpha_2}$ wave states 
has the lowest free energy 
for sufficiently small values of $\epsilon$ 
that do not reverse the signs of the free-energy differences. 
Therefore, 
to summarize the results in this section, 
behavior (c) mentioned in Sect.~\ref{sec:introduction} 
is verified.

\section{\label{sec:effectofdistortion} 
Effect of Lattice Distortion 
}

The effect of the lattice distortion can be 
incorporated by an angle-dependent density of states 
\Equation{eq:rhoxiphi}
{
     \rho(\xi,{\hat \vk}) = 
     \rho_0(\xi,{\hat \vk}) 
     + \epsilon \rho_1(\xi,{\hat \vk}) , 
     }
where 
$\rho_0$ and $\rho_1$ are symmetric and symmetry-breaking parts, 
respectively, 
and 
${\hat \vk} \equiv \vk/|\vk|$. 
For example, 
when 
$(\alpha_1,\alpha_2) = ({\rm d}_{x^2-y^2},{\rm s})$, 
$\rho_0$ and $\rho_1$ are symmetric and antisymmetric 
with respect to the interchange $x \leftrightarrow y$, respectively. 
An example of the symmetry-breaking part 
when $(\alpha_1,\alpha_2) = ({\rm p}_x,{\rm p}_y)$ 
%% is $\rho_1(\xi,{\hat \vk}) \propto \gamma_{{\rm p}_x}({\hat \vk}) 
is $\rho_1 \propto \gamma_{{\rm p}_x}({\hat \vk}) 
                   \gamma_{{\rm p}_y}({\hat \vk})$. 
The summation over $\vk$ is replaced with the integral as 
\Equationnoeqn{eq:suminteg}
{
     \frac{1}{N}\sum_{\vk} \Bigl ( \cdots \Bigr ) 
     = \int \! d \xi \! 
       \int \frac{d^2 {\hat \vk}}{4 \pi}
         \rho(\xi, {\hat \vk}) 
     \Bigl ( \cdots \Bigr ) . 
     }
By using \eq.{eq:rhoxiphi} and the symmetry, 
we obtain $W_{\alpha_1\alpha_2}(\phi = 0) 
= \epsilon {\bar W}_{\alpha_1\alpha_2}$ 
and 
$W_{\alpha_1\alpha_2}^{(2)}(\phi = 0) 
= {\bar W}^{(2)}_{\alpha_1\alpha_2}$, 
where 
\Equationnoeqn{eq:W1W2int}
{
     \begin{split}
     {\bar W}_{\alpha_1\alpha_2}
     & 
     = \int \! d \xi \! 
       \int \frac{d^2 {\hat \vk}}{4 \pi}
         \rho_1(\xi,{\hat \vk}) 
         \left [ 
         W_{\vk}
         \right ]_{\phi = 0} 
         \gamma_{\alpha_1}(\vk)
         \gamma_{\alpha_2}(\vk) , 
     \\ 
     {\bar W}_{\alpha_1\alpha_2}^{(2)} 
     & 
     = \int \! d \xi \! 
       \int \frac{d^2 {\hat \vk}}{4 \pi}
         \rho_0(\xi,{\hat \vk}) 
         \left [ 
         \frac{W_{\vk}}{E_{\vk}^2} 
         \right ]_{\phi = 0} 
         \!\!\!\! \!\! 
         [\gamma_{\alpha_1}(\vk)]^2 
         [\gamma_{\alpha_2}(\vk)]^2 . 
     \end{split}
     }
Hence, \eq.{eq:sinphissol} reduces to 
\Equation{eq:sinphissolepsilon}
{
     \sin \phi = 
     - 
       \frac{ \epsilon {\bar W}_{\alpha_1\alpha_2}}
              {\Delta_{\alpha_1} \Delta_{\alpha_2} 
              {\bar W}_{\alpha_1\alpha_2}^{(2)}} 
     = {\rm O}(\epsilon) . 
     }
Because $|\sin \phi| \leq 1$, 
%% \eqs.{eq:sinphissol} and \refeq{eq:sinphissolepsilon} 
\eq.{eq:sinphissolepsilon} cannot be satisfied 
when $\Delta_{\alpha_1} \Delta_{\alpha_2}$ is too small 
or when $\epsilon$ is too large. 
Therefore, we obtain the upper limit 
of $\epsilon$ 
\Equationnoeqn{eq:epsc2}
{
     \epsilon_{\rm c}^{(2)} 
     = 
       \Delta_{\alpha_1}\Delta_{\alpha_2} 
       {{\bar W}_{\alpha_1\alpha_2}^{(2)}} 
       / 
       {{\bar W}_{\alpha_1\alpha_2}} , 
     }
below which the TRSB state 
is a solution of the gap equation. 
By contrast, 
when an appropriate value of $\phi$ that eliminates 
$\SinGE_{\alpha_1\alpha_2}$ 
does not exist, the TRSB state cannot be a solution, 
and 
the $\alpha_1 \pm \alpha_2$ wave states are only possible solutions. 
$\epsilon_{\rm c}^{(2)}$ 
is not necessarily 
the critical value $\epsilon_{\rm c}$ of $\epsilon$ 
below which the TRSB state is stable, 
because a first-order transition may occur 
at a value of $\epsilon$ smaller than $\epsilon_{\rm c}^{(2)}$. 
If not, a second-order transition occurs at 
$\epsilon = \epsilon_{\rm c}$ ($= \epsilon_{\rm c}^{(2)}$) 
between the TRSB $({\alpha_1} \pm {\alpha_2}) + i {\alpha_2}$ wave state 
and the mixed-symmetry $\alpha_1 \pm \alpha_2$ wave state 
that has the real order parameter 
\Equationnoeqn{eq:alpha1pmalpha2}
{
     \Delta_{\alpha_1} \gamma_{\alpha_1}(\vk) 
     \pm 
     \Delta_{\alpha_2} \gamma_{\alpha_2}(\vk) . 
     }
When we consider the system in which 
$\gamma_{\alpha_1}(\vk)$ has line nodes 
whereas $\gamma_{\alpha_2}(\vk)$ does not have any node, 
the additional real $\alpha_2$-wave component 
shifts the positions of the line nodes 
if $\Delta_{\alpha_2}$ is small. 
On the other hand, if it is large, the nodes vanish.

\section{\label{sec:summary}
Summary and Discussion 
}

To summarize, 
we clarified the manner 
in which a TRSB state becomes stable 
when a distortion breaks the relevant symmetry of a system 
that guarantees 
$S_{\alpha_1 \alpha_2} = 0$.~\cite{NoteScalS} 
The nonzero off-diagonal elements 
$S_{\alpha_1 \alpha_2} = S_{\alpha_2 \alpha_1} \ne 0$ 
destabilize the TRSB states 
including the distorted TRSB 
$(\alpha_1 \pm \alpha_2) + i \alpha_2$ wave states; 
however, 
unless $\Delta_{\alpha_1}\Delta_{\alpha_2}$ is too small, 
the distorted TRSB state can satisfy the gap equation 
by adjusting the shift $\phi$ in the relative phase 
so that $S_{\alpha_1 \alpha_2} = 0$ is restored. 
An analysis of the free-energy functional elucidated that 
in a symmetric system ($\epsilon = 0$), 
if a TRSB state is a solution of the gap function, 
it is at a minimum of the free energy 
and stable against order-parameter fluctuations. 
This implies that when the system is distorted ($\epsilon \ne 0$), 
the TRSB state remains at the minimum 
and stable against the fluctuations 
when the distortion is small. 
We obtained a formula for the upper bound of $\epsilon$ 
below which the TRSB state 
can be a solution of the gap equation.

It is possible that 
the states with and without time-reversal symmetry 
are at local minima of the free energy; 
however, when the TRSB state has the lowest free energy at $\epsilon = 0$, 
small values of $\epsilon$ do not reverse the signs of the differences 
in the free energy. 
Therefore, 
if the TRSB state 
occurs in the symmetric system ($\epsilon = 0$), 
the symmetry-breaking lattice distortion distorts 
the order parameter but does not destabilize the TRSB state 
at sufficiently small values of $\epsilon$.

The present theory does not assume any specific 
$(\alpha_1,\alpha_2)$; hence, 
it can be applied to the TRSB ${\rm d} + i {\rm s}$, 
${\rm p}_x + i {\rm p}_y$, and ${\rm d} + i {\rm d}'$ states. 
We can improve the theory by incorporating the strong coupling effect 
and the effect of the distortion on the coupling constant. 
However, we do not expect that 
these effects significantly change the main part of the present result, 
i.e., 
the finding that the shift in the relative phase restores 
$S_{\alpha_1 \alpha_2} = 0$ 
and the TRSB state remains at a free-energy minimum 
for sufficiently small values of $\epsilon$, 
although the formula for $\epsilon_{\rm c}^{(2)}$ must be modified. 
We leave this task for future research,

\mbox{}

%% JPSJ 
%% \begin{acknowledgments}

%% JPSJ 
%% \end{acknowledgments}

\appendix
\section{
Derivation of \eq.{eq:hatF}
\label{app:A}
}

We use the variational method based on the inequality 
$$
     {\hat F} \equiv \langle H - H_2 \rangle_2 + F_2 \geq F 
     $$ 
for the true free energy $F$ 
and an arbitrary trial Hamiltonian $H_2$, 
where 
$\langle \cdots \rangle_2 = {\rm Tr} [e^{- \beta H_2} \cdots]/Z_2$, 
$F_2 \equiv - \kB T \ln Z_2$, 
and $Z_2 = {\rm Tr}[e^{- \beta H_2}]$. 
We adopt 
\Equationnoeqn{eq:H2}
{
     H_2 = \sum_{\vk \sigma} E_{\vk} 
       \alpha_{\vk \sigma}^{\dagger} 
       \alpha_{\vk \sigma} 
     }
with 
$\alpha_{\vk \sigma} =   u_{\vk} c_{\vk \sigma} 
                       + v_{\vk} c_{- \vk {\bar \sigma}}^{\dagger}$ 
and 
$\alpha_{- \vk {\bar \sigma}}^{\dagger} 
                     = - v_{\vk}^{*} c_{\vk \sigma} 
                       + u_{\vk} c_{- \vk {\bar \sigma}}^{\dagger}$, 
where $u_{\vk}$ (assumed to be real), $v_{\vk}$, 
and $E_{\vk}$ are the variational parameters 
and $u_{\vk}^2 + |v_{\vk}|^2 = 1$ is satisfied. 
We transform the variational parameters $u_{\vk}$ and $v_{\vk}$ 
to $\Delta_{\vk} = |\Delta_{\vk}| e^{i \phi_{\vk}}$ as follows: 
\Equationnoeqn{eq:uvDeltacalE}
{
     \begin{split} 
     u_{\vk} & = 
     \Bigl [ 
     \frac{1}{2} 
     \Bigl ( 1 + \frac{\xi_{\vk}}{{\cal E}_{\vk}} 
     \Bigr ) \Bigr ]^{1/2} , \\ 
     v_{\vk} & = 
     e^{i \phi_{\vk}} 
     \Bigl [ 
     \frac{1}{2} 
     \Bigl ( 1 - \frac{\xi_{\vk}}{{\cal E}_{\vk}} 
     \Bigr ) \Bigr ]^{1/2} , 
     \end{split} 
     }
with ${\cal E}_{\vk} = [ \xi_{\vk}^2 + |\Delta_{\vk}|^2 ]^{1/2}$. 
Hence, ${\hat F}$ 
is a functional of the functions $E_{\vk}$ and $\Delta_{\vk}$. 
A straightforward calculation leads to 
the explicit form given in \eq.{eq:hatF}.

When $E_{\vk}$, $|\Delta_{\vk}|$, and $\phi_{\vk}$ for all $\vk$ 
are independent variational parameters, the extremum conditions 
\Equationnoeqn{eq:BCSextremum}
{
     \pardif{\hat F}{|\Delta_{\vk}|} = 0, ~~~~ 
     \pardif{\hat F}{\phi_{\vk}} = 0, ~~~~ 
     \pardif{\hat F}{E_{\vk}} = 0, 
     }
of ${\hat F}[E,\Delta]$ 
lead to the BCS gap equation 
\Equation{eq:BCSeq}
{
     \Delta_{\vk} =  
       - \frac{1}{N} \sum_{\vk'} V_{\vk\vk'} W_{\vk'} \Delta_{\vk'} , 
     }
with $W_{\vk} 
     = {\tanh(\beta E_{\vk}/2)}/{2 E_{\vk}}$ 
and 
$E_{\vk} = {\cal E}_{\vk} = [\xi_{\vk}^2 + |\Delta_{\vk}|^2 ]^{1/2}$.

If we consider a small fluctuation in a single $\phi_{\vk}$ 
of an arbitrary wave vector $\vk$ 
around the solution of the gap equation, 
it can be proved that 
\Equationnoeqn{eq:d2Fdphik2}
{
     \frac{1}{2N} 
     \frac{\partial^2 {\hat F}}{\partial \phi_{\vk}^2} 
     = W_{\vk} |\Delta_{\vk}|^2 > 0 , 
     }
which implies that any solutions are stable 
against such a simple fluctuation.

\appendix
\section{
Matrix Elements 
\label{app:B}
}

The matrix elements ${\cal K}_{\alpha\alpha'}$ in \eq.{eq:newgapeq} 
are defined as 
\Equationnoeqn{eq:calKdef}
{
     \begin{split}
     {\cal K}_{11} & = {\cal U}_{11} 
     - g_{\alpha_1} {\cal W}_{11}^2 
     - g_{\alpha_2} {\cal W}_{12}^2 , 
     \\ 
     {\cal K}_{22} & = {\cal U}_{22} 
     - g_{\alpha_2} {\cal W}_{22}^2 
     - g_{\alpha_1} {\cal W}_{21}^2 , 
     \\ 
     {\cal K}_{12} 
     = {\cal K}_{21} 
     & = {\cal U}_{12} 
     - (   g_{\alpha_1} {\cal W}_{11} 
         + g_{\alpha_2} {\cal W}_{22} ) {\cal W}_{12} , 
     \end{split}
     }
where 
\Equationnoeqn{eq:calUdef}
{
     {\cal U}_{k \ell} 
      = \frac{1}{N} \sum_{\vk} 
       \frac{E_{\vk}}{{\cal E}_{\vk}} {\cal W}_{\vk} 
       \gamma_{\alpha_k}(\vk) \gamma_{\alpha_{\ell}}(\vk) . 
     }
The matrix elements of ${\hat {\cal L}}$ in \eq.{eq:LmatrixKvector} 
are defined as 
\Equationnoeqn{eq:calLddLssLdmLsmdef}
{
     \begin{split}
     {\cal L}_{\ell \ell} 
     = 
     & \,\, 
     {\cal W}_{\ell \ell} 
        - \Delta_{\alpha_\ell}^2 
        {\cal W}_{\ell \ell\ell \ell}^{(2)} 
           + {\cal L}_{\ell {\bar \ell}} , 
     \\ 
     {\cal L}_{\ell {\bar \ell}} 
     = 
     & \,\, 
     \sin \phi \, \Delta_{\alpha_1} \Delta_{\alpha_2} 
     {\cal W}_{\ell \ell\ell {\bar \ell}}^{(2)} , 
     \\ 
     {\cal L}_{\ell 3} 
     = 
     & \,\, 
       (g_{\alpha_2} \Delta_{\alpha_1}^2 + g_{\alpha_1} \Delta_{\alpha_2}^2) 
       {\cal W}_{\ell \ell\ell {\bar \ell}}^{(2)} 
       - g_{\alpha_{\bar \ell}} {\cal W}_{12} \\
     & ~~~ 
       + \sin \phi \, \Delta_{\alpha_1} \Delta_{\alpha_2} 
         (  g_{\alpha_{\ell}} {\cal W}_{\ell \ell\ell \ell}^{(2)} 
          + g_{\alpha_{\bar \ell}} 
                       {\cal W}_{1122}^{(2)} ) , 
     \\ 
     {\cal L}_{3 \ell} 
     = 
     & \,\, 
       \sin \phi \, \Delta_{\alpha_1} \Delta_{\alpha_2} 
       {\cal W}_{\ell \ell{\bar \ell}{\bar \ell}}^{(2)} 
       - \Delta_{\alpha_\ell}^2 {\cal W}_{\ell \ell\ell {\bar \ell}}^{(2)} 
       + {\cal W}_{12} , 
     \\ 
     {\cal L}_{33} 
     = & \,\, 
       (g_{\alpha_2} \Delta_{\alpha_1}^2 + g_{\alpha_1} \Delta_{\alpha_2}^2) 
       {\cal W}_{1122}^{(2)} 
     \\ 
       & ~~~ 
       + g_{\alpha_2} {\cal L}_{21} + g_{\alpha_1} {\cal L}_{12} - 1 , 
     \end{split}
     }
where $(\ell, {\bar \ell}) = (1,2)$ and $(2,1)$, 
and 
\Equationnoeqn{eq:W2def}
{
     {\cal W}^{(2)}_{ijk\ell}
     = \frac{1}{N} \sum_{\vk} \frac{{\cal W}_{\vk}}{{\cal E}_{\vk}^2}
       \gamma_{\alpha_i}(\vk)
       \gamma_{\alpha_j}(\vk)
       \gamma_{\alpha_k}(\vk)
       \gamma_{\alpha_{\ell}}(\vk) . 
     }

\appendix
\section{
Proof of $\Gamma_0 > 0$ 
\label{app:C}
}

Because \eq.{eq:newgapeq2} 
implies $g_{\alpha} = 1/W_{\alpha \alpha}$ at the extremum point, 
$\Gamma_0$ can be rewritten as 
\Equationnoeqn{eq:Gammaave}
{
    \begin{split}
    \Gamma_0 
    = 
    1 \,\, - \,\, 
          & 
          \Bigl ( 
          \frac{1}
          { \langle [\gamma_{\rm s}(\vk)/\Delta_{\rm d}]^2 \rangle_{W} }
          + 
          \frac{1}
          { \langle [\gamma_{\rm d}(\vk)/\Delta_{\rm s}]^2 \rangle_{W} }
          \Bigr  ) 
    \\ 
    & 
    \times \left \langle 
           \frac{[\gamma_{\rm d}(\vk) \gamma_{\rm s}(\vk)]^2}
                { \xi_{\vk}^2 + [\Delta_{\rm d} \gamma_{\rm d}(\vk)]^2 
                              + [\Delta_{\rm s} \gamma_{\rm s}(\vk)]^2 }
           \right \rangle_{W} , 
    \end{split}
    }
where we defined the average 
\Equation{eq:aveW}
{
     \Bigl \langle \cdots \Bigr \rangle_{W} 
     \equiv 
     \Bigl [ \frac{1}{N} \sum_{\vk} W_{\vk} \Bigr ]^{-1} 
     \frac{1}{N} \sum_{\vk} W_{\vk} (\cdots) . 
     }
It can be proved that 
\Equation{eq:inequality}
{
     \frac{1}{{ 
               \frac{1}{\langle f \rangle} 
              + \frac{1}{\langle g \rangle} }}  
     \geq 
     \left \langle 
     \frac{1}{  { \frac{1}{f} + \frac{1}{g} } }
     \right \rangle 
     }
for arbitrary positive functions $f$ and $g$ of arbitrary variables 
and arbitrary average $\langle \cdots \rangle$ over the variables, 
as shown in Appendix\ref{app:D}. 
The equality sign holds when both $f$ and $g$ are constant. 
By applying this inequality to 
$f \equiv [\gamma_{\rm s}(\vk)/\Delta_{\rm d}]^2$, 
$g \equiv [\gamma_{\rm d}(\vk)/\Delta_{\rm s}]^2$, 
and the average defined by \eq.{eq:aveW} 
and using the fact that $\xi_{\vk}^2 > 0$, 
we obtain $\Gamma_0 > 0$.

\appendix
\section{
Proof of \eq.{eq:inequality}
\label{app:D}
}

It is sufficient to prove \eq.{eq:inequality} 
with respect to the simple average 
$$
     \langle x \rangle = \frac{1}{n} \sum_{k=1}^{n} x_k 
     $$
because any average with any weight can be written in this form 
with a sufficiently large $n$. 
All the $x_k$ and $y_k$ are assumed to be positive. 
Equation~\refeq{eq:inequality} holds 
for $n = 2$ because 
\Equation{eq:N2}
{
     \frac{(x_1 + x_2)(y_1 + y_2)}{x_1 + x_2 + y_1 + y_2} 
     \geq 
     \frac{x_1 y_1}{x_1 + y_1} 
     + 
     \frac{x_2 y_2}{x_2 + y_2} 
     }
for positive $x_k$ and $y_k$. 
When \eq.{eq:inequality} holds for $n$, 
it is verified that it holds for $n + 1$. 
Indeed, with definitions 
$X \equiv \sum_{k = 1}^{N} x_k$, 
$Y \equiv \sum_{k = 1}^{N} y_k$, 
$x \equiv x_{N + 1}$, and 
$y \equiv y_{N+1}$, 
the induction hypothesis is written as 
\Equationnoeqn{eq:forN}
{
     \frac{XY}{X + Y} - \sum_{k = 1}^{n} \frac{x_ky_k}{x_k + y_k}
     \geq 0 , 
     }
and hence, it follows that 
\Equationnoeqn{eq:Nplus1}
{
     \begin{split}
     & \frac{(X + x)(Y + y)}{X + x + Y + y} 
     - \sum_{k = 1}^{n + 1} \frac{x_k y_k}{x_k + y_k} 
     \\ 
     & \geq 
     \frac{(X + x)(Y + y)}{X + Y + x + y} 
     - 
     \frac{XY}{X + Y} 
     - 
     \frac{xy}{x + y} 
     \geq 0 , 
     \end{split}
     }
where the last inequality follows from \eq.{eq:N2} 
with $x_1 = X$, $x_2 = x$, $y_1 = Y$, and $y_2 = y$. 
Therefore, we obtain 
\Equationnoeqn{eq:inequality_xy}
{
     \frac{1}{{ 
               \frac{1}{\langle x \rangle} 
              + \frac{1}{\langle y \rangle} }}  
     \geq 
     \left \langle 
     \frac{1}{  { \frac{1}{x} + \frac{1}{y} } }
     \right \rangle 
     }
by mathematical induction. 
The equality sign holds when neither $x_k$ nor $y_k$ depends on $k$. 
Therefore, \eq.{eq:inequality} has been proved. 
The inequation can be easily extended as 
$$
     \frac{1}{\sum_{\ell} \frac{1}{\langle f^{(\ell)} \rangle }} 
     \geq 
     \left \langle 
     \frac{1}{\sum_{\ell} \frac{1}{f^{(\ell)}}} 
     \right \rangle 
     $$
for positive functions $f^{(\ell)}$.

%%%%%%%%%%%%%%%%%%%%%%%%%%%%%%%%%%%%%%%%%%%%%%%%%%%%%%%%%%%%%%%%%%%%%%%
%%  References                                                       %%
%%%%%%%%%%%%%%%%%%%%%%%%%%%%%%%%%%%%%%%%%%%%%%%%%%%%%%%%%%%%%%%%%%%%%%%

%%%%%%%%%%%%%%%%%%%%%%%%%%%%%%%%%%%%%%%%%%%%%%%%%%%%%%%%%%%%%%%%%%%%%%%


\begin{thebibliography}{99}
\bibitem{NoteSRONMR}  % Sr2RuO4, note on NMR
  In ${\rm Sr_2RuO_4}$, 
  a recent study showed a pronounced drop in 
  $^{17}{\rm O}$ NMR Knight shift, 
  indicating that 
  the superconducting state 
  has a component of antiparallel spin pairing.~\cite{Pus19,Ish20} 
  The temperature dependence of the upper critical field 
  also suggests that such a component occupies a major part 
  in the superconducting state. 
  On the other hand, 
  the TRSB state is supported by a measurement of 
  muon spin relaxation.~\cite{Luk98} 
  These behaviors can be consistently explained 
  if the superconducting state is 
  an interlayer chiral singlet state 
  with an order parameter of the form 
  $$
       \Delta(\vk) \propto 
          \sin k_z \, [\gamma_{{\rm p}_x}(\vk_{\parallel}) 
               + i \gamma_{{\rm p}_y}(\vk_{\parallel})] , 
  $$
  where $\vk_{\parallel} = (k_x,k_y)$. 
  In accordance with experimental observations,~\cite{Kit10,Kit18} 
  the transition temperature of this state 
    is unaffected by perpendicular pressures 
  and 
  the order parameter has a horizontal line node.~\cite{Shi20c}
\bibitem{NoteScalS}
  Whether the off-diagonal elements 
    ${\bar S}_{\alpha_1\alpha_2}$ and 
    ${\bar S}_{\alpha_2\alpha_1}$ are zero 
    is practically equivalent to 
    whether the off-diagonal elements 
    $S_{\alpha_1\alpha_2} = S_{\alpha_2\alpha_1}$ 
    of the scattering matrix ${\hat S}$ are zero. 
  The scattering matrix is written in the form 
  $$
       {\hat S} = {\hat V} + {\hat V} {\hat W}{\hat V} + \cdots 
       = (1 - {\hat \SinGE})^{-1} {\hat V} , 
  $$
  with ${\hat \SinGE} \equiv {\hat V}{\hat W}$, 
  where 
  ${\hat V}$ and ${\hat W}$ are matrices having elements 
  $V_{\vk \vk'}$ and $W_{\vk} \delta_{\vk \vk'}$, respectively. 
  Here, 
  \Equationnoeqn{eq:WGG}
  {
       W_{\vk} = - T \sum_{n} 
       G_{\sigma}(\vk,i\omega_n) 
       G_{{\bar \sigma}}(-\vk,-i\omega_n) , 
       }
  where $G_{\sigma}(\vk,i\omega_n)$ is the Green's function of electrons. 
  The matrix elements $\SinGE_{\alpha \alpha'}$ in the text 
  are expressed as 
  \Equationnoeqn{eq:S12}
  {
       \SinGE_{\alpha \alpha'}
         = \frac{1}{N^2} \sum_{\vk} \sum_{\vk'} 
             \gamma_{\alpha}(\vk) 
                \SinGE_{\vk\vk'}
             \gamma_{\alpha'}(\vk') , 
       }
  where $\SinGE_{\vk\vk'}$ is the element of ${\hat \SinGE}$ 
  and $N$ denotes the number of lattice sites. 
  Similarly, we can reduce the matrices ${\hat S}$, 
  ${\hat V}$, and ${\hat W}$ 
  to $2 \times 2$ matrices in the subspace spanned by 
  $\gamma_{\alpha_1}$ and $\gamma_{\alpha_2}$. 
  In this subspace, ${\hat S}$ can have nonzero off-diagonal elements 
  only when at least one of ${\hat V}$ and ${\hat W}$ is not diagonal. 
  Hence, 
  whether ${\hat \SinGE}$ is diagonal is practically equivalent to 
  whether ${\hat S}$ is diagonal. 
  The only exception is the case in which 
  ${\hat \SinGE} = {\hat V} {\hat W}$ is diagonal 
  even when neither ${\hat V}$ nor ${\hat W}$ are diagonal; 
  however, such a case can occur only accidentally.
\bibitem{NoteSym}
  This form cannot be excluded merely because 
  the symmetries $\alpha_1$ and $\alpha_2$ 
  belong to the same irreducible representation 
  in distorted systems.
\bibitem{NoteGnrl}
  The form of $\Delta_{\vk}$ in \eq.{eq:chiralphi} 
  is generic as long as 
  the coupling constant of the pairing interaction has the form 
  $$
       V_{\vk \vk'} = - \sum_{\ell = 1,2} \sum_{\ell' = 1,2} 
         g_{\alpha_{\ell} \alpha_{\ell'}} 
           \gamma_{\alpha_{\ell}}(\vk) 
           \gamma_{\alpha_{\ell'}}(\vk') . 
       $$
  Furthermore, 
  the fact that 
  the imaginary part is a pure $\alpha_2$-wave function 
  in \eq.{eq:chiralphiReIm} 
  does not affect the generality. 
  Because the arbitrary overall phase factor $e^{i \theta}$ 
  can be multiplied, 
  \eq.{eq:chiralphi} includes an arbitrary 
  $\alpha_1' \pm i \alpha_2'$ wave state, 
  where $\gamma_{\alpha_1'}$ and $\gamma_{\alpha_2'}$ 
  are arbitrary linear combinations of 
  $\gamma_{\alpha_1}$ and $\gamma_{\alpha_2}$.
\bibitem{NoteDiag}
  It may be considered that because the $2 \times 2$ matrix 
  $$
       {\hat \SinGE}_2 \equiv 
       \left ( 
       \begin{array}{cc}
       {\bar S}_{\alpha_1\alpha_1} & {\bar S}_{\alpha_1\alpha_2} \\
       {\bar S}_{\alpha_2\alpha_1} & {\bar S}_{\alpha_2\alpha_2} 
       \end{array}
       \right ) 
       $$
  can be diagonalized by appropriate basis functions 
  $\gamma_{\alpha_1'}$ and $\gamma_{\alpha_2'}$, 
  which are linear combinations of 
  $\gamma_{\alpha_1}$ and $\gamma_{\alpha_2}$, 
  the TRSB state can be easily constructed in terms of 
  $\gamma_{\alpha_1'}$ and $\gamma_{\alpha_2'}$ 
  as 
  $$
       \Delta_{\vk} = \Delta_{\alpha_1'} \gamma_{\alpha_1'}(\vk) 
         + i \Delta_{\alpha_2'} \gamma_{\alpha_2'}(\vk) . 
       $$
  This state may appear to avoid the destabilization by 
  the off-diagonal scattering 
  because 
    $\SinGE_{\alpha_1'\alpha_2'} = \SinGE_{\alpha_2'\alpha_1'} = 0$. 
  However, it can be easily verified that 
  unless $\SinGE_{\alpha_1\alpha_2} = \SinGE_{\alpha_2\alpha_1} = 0$, 
  although $\SinGE_{\alpha_1'\alpha_2'} = \SinGE_{\alpha_2'\alpha_1'} = 0$, 
  either $\Delta_{\alpha_1'}$ or $\Delta_{\alpha_2'}$ is equal to zero, 
  which implies that this state is not a TRSB state. 
  The diagonalization is a mathematically equivalent transformation 
  of the gap equation, 
  which does not affect the fact that 
  the nonzero off-diagonal elements $\SinGE_{\alpha_1\alpha_2}$ 
  and $\SinGE_{\alpha_2\alpha_1}$ destabilize the TRSB states. 
  Furthermore, $\Delta_{\vk}$ of this form is 
  included in $\Delta_{\vk}$ in \eq.{eq:chiralphi}
  because the overall phase factor is arbitrary;~\cite{NoteGnrl} 
  hence, 
  the problem about the stability of the TRSB state with this $\Delta_{\vk}$ 
  in terms of $\gamma_{\alpha_{\ell}'}$ 
  is included 
  in the original problem about 
  the stability of the TRSB states expressed in \eq.{eq:chiralphi}.
\bibitem{NoteSRO}  % 
  For the compound ${\rm Sr_2RuO_4}$, 
  the effects of uniaxial strains 
  in the $[1,0,0]$, $[0,1,0]$, and $[0,0,1]$ directions 
  were experimentally examined.~\cite{Hic14,Kit10}
\bibitem{NoteJur99}
  In Ref.~\citen{Jur99}, 
  $\Delta(\varphi) = \Delta_{\rm 0} + i \Delta_{\rm 1} \sqrt{2} \cos 2 \varphi$ 
  is rewritten 
  as $\Delta(\phi) = \Delta_{\rm 0}' + i \Delta_{\rm 1}' F(\phi)$, 
  where $F(\phi) = f(\phi) - {\bar f}$, 
  $f(\phi) = \sqrt{2} \cos 2 \varphi$, 
  ${\bar f} = \langle f(\phi) \rangle_{\phi}$, 
  $\Delta_{\rm 0}' = \Delta_{\rm 0} + i {\bar f} \Delta_{\rm 1}$, 
  $\Delta_{\rm 1}' = \Delta_{\rm 1}$, 
  and $\phi$ is defined by $\tan \phi = (m_a/m_b)^{1/2} \tan \varphi$. 
  The authors called it 
  the ${\rm s} + i ({\rm s+d})$-symmetric state; 
  however, 
  it is different from 
  the present distorted TRSB state 
  because the above transformation of $\Delta(\phi)$ is an identity 
  and the relative phase between the two terms 
  $\Delta_{\rm 0}$ and $i \Delta_{\rm 1} \sqrt{2} \cos 2 \varphi$ 
  remains $\pi/2$.
\bibitem{NoteOrth}
  It seems that in most papers 
  that examine ${\rm d} + i{\rm s}$ states 
  in orthorhombic systems,~\cite{Jur99,Liu97,Gho99} 
  the off-diagonal elements $\SinGE_{\rm sd}$ and $\SinGE_{\rm ds}$ 
  are incorrectly treated. 
  For example, 
  it seems that 
  $\SinGE_{\rm sd}$ and $\SinGE_{\rm ds}$ are simply ignored 
  or incorrectly considered to be zero in Refs.~\citen{Liu97} and \citen{Gho99}. 
  The authors examined systems in which 
  $$
       V_{\vk \vk'} = - g_{\rm s} \gamma_{\rm s}(\vk) \gamma_{\rm s}(\vk') 
                      - g_{\rm d} {\tilde \gamma}_{\rm d}(\vk) 
                                  {\tilde \gamma}_{\rm d}(\vk') 
       $$ 
  and $\xi_{\vk} = - 2 t (\cos k_x + \beta \cos k_y) 
                    + 4 t' \cos k_x \cos k_y - \mu$, 
  and they assumed the following form of the gap function 
  $\Delta_{\vk} = \Delta_{\rm s} \gamma_{\rm s}(\vk) 
    + i \Delta_{\rm d} {\tilde \gamma}_{\rm d}(\vk)$, 
  where $\gamma_{\rm s}(\vk) = 1$ and 
  ${\tilde \gamma}_{\rm d}(\vk) = \cos k_x - \beta' \cos k_y$ 
  with $\beta \ne 1$ and $\beta' \ne 1$ 
  due to orthorhombic distortion. 
  By substituting these expressions into the BCS gap equation, 
  we obtain the real part 
  $$\Delta_{\rm s} \gamma_{\rm s}(\vk) 
   = \gamma_{\rm s}(\vk) \SinGE_{\rm ss} \Delta_{\rm s} 
     + {\tilde \gamma}_{\rm d}(\vk) \SinGE_{\rm ds} \Delta_{\rm s}$$
  and the imaginary part 
  $$ \Delta_{\rm d} {\tilde \gamma}_{\rm d}(\vk) 
   =    \gamma_{\rm s}(\vk) \SinGE_{\rm sd} \Delta_{\rm d} 
     +  {\tilde \gamma}_{\rm d}(\vk) \SinGE_{\rm dd} \Delta_{\rm d}, 
  $$
  with 
  \Equationnoeqn{eq:LiuGho}
  {
       \begin{split} 
       \SinGE_{\rm ss} & = g_{\rm s} 
                      \int \frac{d^2 \vk}{(2 \pi)^2}
                      \frac{\tanh (E_{\vk}/2T)}{2E_{\vk}},  \\ 
       \SinGE_{\rm dd} & = g_{\rm d} 
                      \int \frac{d^2 \vk}{(2 \pi)^2}
                      \frac{\tanh (E_{\vk}/2T)}{2E_{\vk}} 
                      [{\tilde \gamma}_{\rm d}(\vk)]^2 ,  \\ 
       \SinGE_{\rm sd} & = g_{\rm s} 
                      \int \frac{d^2 \vk}{(2 \pi)^2}
                      \frac{\tanh (E_{\vk}/2T)}{2E_{\vk}} 
                      {\tilde \gamma}_{\rm d}(\vk) , \\ 
       \SinGE_{\rm ds} & = g_{\rm d} 
                      \int \frac{d^2 \vk}{(2 \pi)^2}
                      \frac{\tanh (E_{\vk}/2T)}{2E_{\vk}} 
                     {\tilde \gamma}_{\rm d}(\vk) , 
       \end{split}
       } 
  where $E_{\vk} = \sqrt{\xi_{\vk}^2 + |\Delta_{\vk}|^2}$. 
  Here, we should note that in their theory, 
  $E_{\vk}$ does not include the relative phase $\phi$; 
  hence, in general, $\SinGE_{\rm sd} \ne 0$ and $\SinGE_{\rm ds} \ne 0$ 
  when $\beta \ne 1$ and/or $\beta' \ne 1$, 
  except for accidental values of model parameters such as 
  $t' = \mu = 0$. 
  Note that $\xi_{\vk}^2 \ne \xi_{\vk + \vQ}^2$ 
  unless $t' = \mu = 0$, although $|\Delta_{\vk}|^2 = |\Delta_{\vk + \vQ}|^2$, 
  where $\vQ = (\pi,\pi)$. 
  Because $\gamma_{\rm s}$ and ${\tilde \gamma}_{\rm d}$ 
  are linearly independent, 
  these equations lead to 
  $\Delta_{\rm s} = \SinGE_{\rm ss} \Delta_{\rm s}$, 
  $\Delta_{\rm d} = \SinGE_{\rm dd} \Delta_{\rm d}$, 
  and 
  $0 = \SinGE_{\rm ds} \Delta_{\rm s} 
     = \SinGE_{\rm sd} \Delta_{\rm d}$. 
  However, the authors retained only the former two equations 
  and dropped the last two equations 
  even when $\SinGE_{\rm sd} \ne 0$ and $\SinGE_{\rm ds} \ne 0$.
\bibitem{NotedetL}
  In the model given in \eq.{eq:rhoxiphi}, 
  it can be verified that $\det[{\cal L}] \ne 0$ 
  by examining factors of the power functions $\epsilon^{k}$.
\end{thebibliography}
\end{document}